\documentstyle[aps,twocolumn,epsf,psfig,amssymb,amsmath,floats]{revtex}

\def\s0#1#2{\mbox{\small{$ \frac{#1}{#2} $}}}
\def\0#1#2{\frac{#1}{#2}}

\begin{document}
\twocolumn[\hsize\textwidth\columnwidth\hsize\csname
@twocolumnfalse\endcsname
\title{Illuminating Dense Quark Matter}
\author{
\hfill 
Cristina Manuel$^a$
and
Krishna Rajagopal$^b$
\hfill\raisebox{21mm}[0mm][0mm]{\makebox[0mm][r]{ 
CERN-TH-2001-191, MIT-CTP-3165}}%
}
\address{$^a$Theory Division, CERN, CH-1211 Geneva 23, Switzerland.\\ 
$^b$Center for Theoretical Physics, Massachusetts Institute of Technology,
Cambridge, MA 02139, USA }
\maketitle

\begin{abstract}\noindent
{We imagine shining light on a lump of cold dense quark
matter, in the CFL phase and therefore a transparent insulator.
We calculate the angles of reflection and refraction, and the intensity of
the reflected and refracted light.  Although the only
potentially observable
context for this phenomenon (reflection of light
from and refraction of light through
an illuminated quark star) is unlikely to be realized,   
our calculation casts new light on the old idea that
confinement makes the QCD vacuum behave as if filled with a
condensate of color-magnetic monopoles.
\\

PACS numbers: 
}

\end{abstract}
\vskip2.pc]

At high enough baryon density and low temperature,
the ground state of QCD with three flavors of quarks is the
color-flavor locked (CFL) phase~\cite{CFL,reviews}.
In this phase, quarks of all three colors and all three flavors
form Cooper pairs, meaning that all fermionic quasiparticles
are gapped. The gap $\Delta$ is likely of order tens 
to 100 MeV at astrophysically accessible densities,
with quark chemical potential $\mu\sim(350-500)$ MeV~\cite{reviews}. 
The condensate is
charged with respect to eight of the nine massless gauge bosons 
(eight gluons, one photon) of the
ordinary vacuum, meaning that eight gauge bosons get a mass.
Chiral symmetry is spontaneously
broken, and so is baryon number   (i.e., the
material is a superfluid.)
At asymptotic densities, 
the effective coupling is 
weak and the properties of the 
ground state and its low-energy excitations can
be determined quantitatively
by adapting methods used in the theory
of superconductivity (BCS theory).

The CFL phase persists for finite masses and even for
unequal masses, so long as the differences are not too 
large~\cite{ABR2+1}.  It is very likely the ground
state for real QCD, assumed to be in equilibrium
with respect to the weak interactions, over a substantial
range of densities.  Throughout the range of parameters
over which the CFL phase exists as a bulk (and therefore
electrically neutral) phase, it consists of equal
numbers of $u$, $d$ and $s$ quarks and is therefore
electrically neutral in the absence of any 
electrons \cite{neutrality}.    
The equality of the three quark number densities is 
enforced in the CFL phase
by the fact that this equality maximizes the pairing energy
associated with the formation of $ud$, $us$, and $ds$ Cooper pairs.
This equality is enforced even though the strange quark, with mass $m_s$,
is heavier than the light quarks~\cite{BedaqueSchaefer}.

In the CFL phase, there is an unbroken $U(1)_{\tilde Q}$ gauge symmetry
and a corresponding massless
photon given by a linear combination of the 
ordinary photon and one of the gluons~\cite{CFL,ABRflux,reviews}.  
$U(1)_{\tilde Q}$ is generated by
$\tilde Q = Q + T_8/\sqrt{3}$, 
where $Q$ is the conventional electromagnetic
charge generator and the color hypercharge generator
$T_8$ is normalized 
such that,
in the representation of the quarks,
$T_8/\sqrt{3}=\mbox{diag}(-\s0{2}{3},\s0{1}{3},\s0{1}{3})$  
in color space.
The CFL condensate
is $\tilde Q$-neutral, the $U(1)$ symmetry generated
by $\tilde Q$
is therefore unbroken, the associated $\tilde Q$-photon remains massless,
and within the CFL phase the $\tilde Q$-electric and $\tilde Q$-magnetic
fields satisfy Maxwell's equations. The massless 
combination of the photon and the eighth gluon,
$A_\mu^{\tilde Q}$, and
the orthogonal massive combination which
experiences the Meissner effect, $A_\mu^X$,
are given by
\begin{eqnarray}
A^{\tilde Q}_\mu &  = &  \cos\theta A_\mu + \sin\theta G^8_\mu \ , 
\label{AmuQtilde}\\
A^X_\mu  & = &  -\sin\theta A_\mu + \cos\theta G^8_\mu\ .
\end{eqnarray}
The
mixing angle $\theta$ (called $\alpha_0$ in
Ref. \cite{ABRflux})
which specifies the unbroken $U(1)$ is given by
\begin{equation}\label{rot:alpha0}
\cos\theta = \frac{g}{\sqrt{ g^2 + e^2/3}}\ .
\end{equation}
$\theta$ is the analogue of the Weinberg
angle in electroweak theory.
At accessible densities, the gluons are strongly coupled
($g^2/4\pi \sim 1$)
and the photons are weakly coupled
($e^2/4\pi \approx 1/137$), so $\theta$ 
is small, perhaps of order $1/20$.
The ``rotated photon''
consists mostly of the usual photon, with only a small
admixture of the $G^8$ gluon.

All elementary excitations in the CFL phase 
are either $\tilde Q$-neutral or 
couple to $A^{\tilde Q}_\mu$
with charges which
are integer multiples of the $\tilde Q$-charge of the electron
$\tilde e ~=~ e\cos\theta$,
which is less than $e$ because the electron couples only
to the $A_\mu$ component of $A^{\tilde Q}_\mu$. 
The only
massless excitation (the superfluid mode) is $\tilde Q$-neutral.
Because all charged excitations have nonzero mass and there
are no electrons present,
the CFL phase in bulk is a transparent insulator at
low temperatures:
$\tilde Q$-magnetic and $\tilde Q$-electric fields 
within it evolve 
simply according to Maxwell's equations, and low frequency $\tilde Q$-light
traverses it without scattering.

Imagine shining light 
on a chunk of
dense quark matter in the CFL phase.  
If CFL matter occurs only within the cores of
neutron stars, cloaked under kilometers of hadronic
matter~\cite{interface}, the thought experiment we describe here in which 
light waves travelling in vacuum strike CFL matter can never
arise in nature.  If, however, the fact that quark matter
features many more strange quarks than 
ordinary nuclear matter renders it stable 
even at zero pressure,
then one may imagine quark stars in nature~\cite{Witten}.  
Such a quark star may be made of CFL quark matter
throughout, or may have an outer layer in which a less
symmetric pattern of pairing occurs. For example, quarks
of only two colors and flavors may pair, yielding the 2SC 
phase which was the first color superconducting phase
studied \cite{2SC}. Some of the remaining quarks with differing Fermi
momenta may also form a crystalline color superconductor~\cite{crystalline}.
As in the CFL phase, the 2SC condensate leaves a (slightly different) 
${\tilde Q}$-photon massless. However,
the 2SC phase is a good $\tilde Q$-conductor because of
the presence of  
unpaired quarks and electrons.  Thus, 2SC matter is opaque and metallic
rather than transparent and insulating.   Illuminating it would
result in absorption and reflection, but no refraction. 
We shall assume that the quark matter we illuminate is
in the transparent CFL phase all the way out to its surface.

Consider, then,
an enormously dense, but transparent, illuminated quark star.
Some light falling on its surface will reflect,
and some will refract into the star in the form
of $\tilde Q$-light.  We shall
calculate the reflection and refraction angles and the intensity
of the reflected light and refracted $\tilde Q$-light.  
The partial Meissner effect induced by
a static magnetic field has been
analyzed previously~\cite{ABRflux}.  We analyze
a time-varying electromagnetic field. As a bonus, 
our analysis allows us to use
well understood properties of
dense quark matter in the CFL phase to learn 
about the (less well understood) QCD vacuum.

We assume that the light has $\omega$ and $k$ both much less
than the energy needed to create a charged excitation in the CFL phase. 
This means $\omega,k \ll \Delta$, where $\Delta$ is the fermionic gap,
to avoid the breaking of pairs
and the creation of quasiparticles. It also means 
$\omega,k \ll m_{\pi^\pm},m_{K^\pm}$, 
where $\pi^\pm$ and $K^\pm$ are the charged 
pions and kaons of 
the CFL phase. Their masses are of order 
${\tilde e} \Delta$ in the chiral limit~\cite{Manuel:2001xt},
and the contribution to their masses 
due to finite quark masses has also
been evaluated~\cite{MesonMasses}.

In vacuum the electromagnetic fields obey the free Maxwell's
equations
\begin{mathletters}
\label{Maxwell}
\begin{eqnarray}
{\bf \nabla} \cdot {\bf D} & = & 0 \ , \qquad {\bf \nabla} \times {\bf E} =
-  \frac{\partial {\bf B}}{\partial t} \ , \\
{\bf \nabla} \cdot {\bf B} & = & 0 \ , \qquad {\bf \nabla} \times {\bf H} =
 \frac{\partial {\bf D}}{\partial t} \ ,
\end{eqnarray}
\end{mathletters}
where ${\bf D} = \epsilon_0 {\bf E}$ and ${\bf B} = \mu_0 {\bf H}$,
and $\epsilon_0$ and $\mu_0$ are the vacuum dielectric constant and magnetic
permeability, respectively, such that the velocity of light $c =1/\sqrt{\mu_0
\epsilon_0}$. Deep in the CFL phase, the rotated fields
${\tilde {\bf E}}$ and ${\tilde {\bf B}}$  obey the same  field equations,
but with dielectric constant~\cite{Litim:2001mv}
\begin{equation}
\label{rot-dielct}
{\tilde \epsilon} = \epsilon_0 \left(1 + \frac{ 8{\alpha} }{9 \pi} \cos^2{\theta}
\frac{\mu^2}{\Delta^2}\right) \ ,
\end{equation}
where  $\alpha$ is the 
electromagnetic fine structure constant and $\mu$ is
the chemical potential.  This expression for $\tilde\epsilon$
is valid to leading order in $\alpha$, and for $\omega, k \ll \Delta$.
The dependence of $\tilde\epsilon$ on $\omega$ arises only in corrections
to (\ref{rot-dielct}) which are suppressed by $\omega^2/\Delta^2$,
and we therefore neglect dispersion in this letter.
The magnetic permeability
in the CFL phase remains unchanged to leading order,  ${\tilde \mu} = \mu_0$.
The index of refraction of CFL quark matter thus reduces to
${\tilde n} = \sqrt{{\tilde \mu} {\tilde \epsilon}/\mu_0\epsilon_0 } =
\sqrt{{\tilde \epsilon}/\epsilon_0 }$. If we apply (\ref{rot-dielct})
for $\mu/\Delta\sim (4-10)$, we obtain ${\tilde n} \sim (1.02 -1.1)$.

We take the surface of the CFL matter to be
planar, with the CFL phase at $z>0$ and vacuum at $z<0$.
(That is, we assume any curvature of the surface
is on length scales long compared to the wavelength of the light.)
For an ordinary dielectric, the analogous problem is
solved in Ref. \cite{Jackson}.
The complication here is that we must match the ordinary
electric and magnetic fields in vacuum onto $\tilde Q$-electric
and $\tilde Q$-magnetic fields within the CFL phase.
The properties of the reflected and refracted waves
will therefore depend upon both the dielectric
constant $\tilde \epsilon$ and the mixing angle $\theta$.

We are only interested in the reflected and refracted waves, and
not in the detailed field configurations very close
to the interface. This means that we can follow
the strategy of Ref. \cite{ABRflux} and encapsulate
the physics of the interface into boundary conditions
relating $\bf E$ and $\bf B$ on the vacuum side of
the interface to ${\tilde{\bf E}}$ and ${\tilde{\bf B}}$
on the CFL side.  On the CFL side,
the massive $X$ fields can be neglected as long as $z$ is
greater than some $\lambda^{\rm CFL}$,
while on the vacuum side, the confined gluon fields can be
neglected as long as $|z|$ is greater than some  $\lambda^{\rm QCD}$.
$\lambda^{\rm QCD}$
is a length scale characteristic of confinement. For the non-static
fields of interest, and in the weak coupling regime, $\lambda^{\rm CFL}$ 
is of order $1/\Delta$, longer than the inverse Meissner
mass $\sim g\mu$~\cite{Rischke:2000ra}.  
In order to describe light whose wavelength is long compared
to  $\lambda^{\rm QCD}$ and 
$\lambda^{\rm CFL}$, we need boundary conditions
relating $\bf E$ and $\bf B$ at $z=-\lambda^{\rm QCD}$ to
${\tilde{\bf E}}$ and ${\tilde{\bf B}}$ at $z=+\lambda^{\rm CFL}$.

$X$-magnetic fields experience a Meissner effect in the CFL phase, meaning 
that supercurrents 
in the CFL matter 
within $\lambda^{\rm CFL}$ 
of the interface screen the
$X$-component of any ordinary magnetic field parallel 
to the interface on the vacuum
side, yielding
the boundary condition
\begin{equation} 
{\tilde {\bf H}}_{\parallel} (t,x,y,\lambda^{\rm CFL}) =  \cos{\theta} \,
{\bf H}_{\parallel} (t,x,y,-\lambda^{\rm QCD})\ .
\label{ParallelH}
\end{equation}
The CFL condensate is charged with respect to the $X$ gauge boson,
meaning that if there is
an ordinary electric field perpendicular to the interface on
the vacuum side, the $X$ component
of the electric flux will terminate in the CFL phase within
$\lambda^{\rm CFL}$ 
of the interface, yielding
\begin{equation}
{\tilde {\bf D}}_{\perp} (t,x,y,\lambda^{\rm CFL}) =  \cos{\theta} \,
{ \bf D}_{\perp} (t,x,y,-\lambda^{\rm QCD}) \ .
\label{PerpendicularD}
\end{equation}
We expect that the confined QCD vacuum should behave as if
it is a condensate of color-magnetic monopoles \cite{tHooft}.  
That is, in the vacuum
color magnetic field lines end: if there is 
a $\tilde Q$-magnetic field perpendicular to the interface
on the CFL side, the vacuum will ensure that only the ordinary
magnetic field is admitted. Thus,
\begin{equation}
{\bf B}_{\perp} (t,x,y,-\lambda^{\rm QCD}) =
\cos\theta \, {\tilde {\bf B}}_{\perp} (t,x,y,\lambda^{\rm CFL}) 
\ .
\label{PerpendicularB}
\end{equation}
Finally, color magnetic currents on the vacuum side of 
the interface should exclude the color component of any
$\tilde Q$-electric field parallel to the interface on
the CFL side, ensuring that
\begin{equation}
{ \bf E}_{\parallel} (t,x,y,-\lambda^{\rm QCD})=
\cos\theta \, {\tilde {\bf E}}_{\parallel} (t,x,y,\lambda^{\rm CFL}) 
 \ .
\label{ParallelE}
\end{equation}
At sufficiently high density, the property of CFL matter from which
(\ref{ParallelH}) and (\ref{PerpendicularD}) follow,
namely the Meissner effect for $X$-bosons, is a weak-coupling
phenomenon which can
be understood analytically.
The properties of the QCD vacuum used to deduce
(\ref{PerpendicularB}) and (\ref{ParallelE}) follow from a  
reasonable and familiar description
of confinement as a dual Meissner effect, 
but confinement is not yet understood analytically.
It is therefore of interest that our analysis below provides
a {\it derivation} of (\ref{PerpendicularB}) and (\ref{ParallelE}) 
from (\ref{ParallelH}).

Consider an incident wave, with wave vector
${\bf k} = \frac{\omega}{c} (\sin{i}, 0, \cos{i})$, so that
\begin{equation}
\label{incident}
{\bf E} = \boldsymbol{\mathcal {E}} e^{i({\bf k}\cdot {\bf r} - \omega t)} \ ,
\qquad {\bf B} = \sqrt{\mu_0 \epsilon_0}\, \frac{{\bf k}}{k} \times {\bf E} \ .
\end{equation}
There are two orthogonal linear polarizations,
shown in Fig. 1, which we will treat separately.
In the first, the vector $\boldsymbol{\mathcal {E}}$ 
is parallel to the interface
while in the second, $\boldsymbol{\mathcal {E}}$ 
lies in the plane of incidence.
The reflected wave is 
\begin{equation}
\label{reflected}
{\bf E}' = \boldsymbol{\mathcal {E}}' 
e^{i({\bf k'}\cdot {\bf r} - \omega t)} \ ,
\qquad {\bf B'} = \sqrt{\mu_0 \epsilon_0}\, \frac{{\bf k'}}{k'} 
\times {\bf E'} \ ,
\end{equation}
with wave vector ${\bf k}' = \frac{\omega}{c} (\sin{i'} , 0, -\cos{i'})$,
while the refracted wave is
\begin{equation}
\label{refracted}
{\tilde {\bf E}}_r = \tilde{\boldsymbol{\mathcal {E}} }_r 
e^{i({\bf k_r}\cdot {\bf r} - \omega t)} \ ,
\qquad {\tilde {\bf B}}_r = \sqrt{{\tilde \mu}  {\tilde \epsilon}}
\, \frac{{\bf k_r}}{k_r} \times {\tilde {\bf E}}_r \ ,
\end{equation}
with wave vector ${\bf k_r} = \omega \, \sqrt{{\tilde \mu}  {\tilde \epsilon}}  (\sin{r}, 0, \cos{r})$.

\begin{figure}[t]
\epsfxsize=2.7in
\hspace*{0.26in}
\epsffile{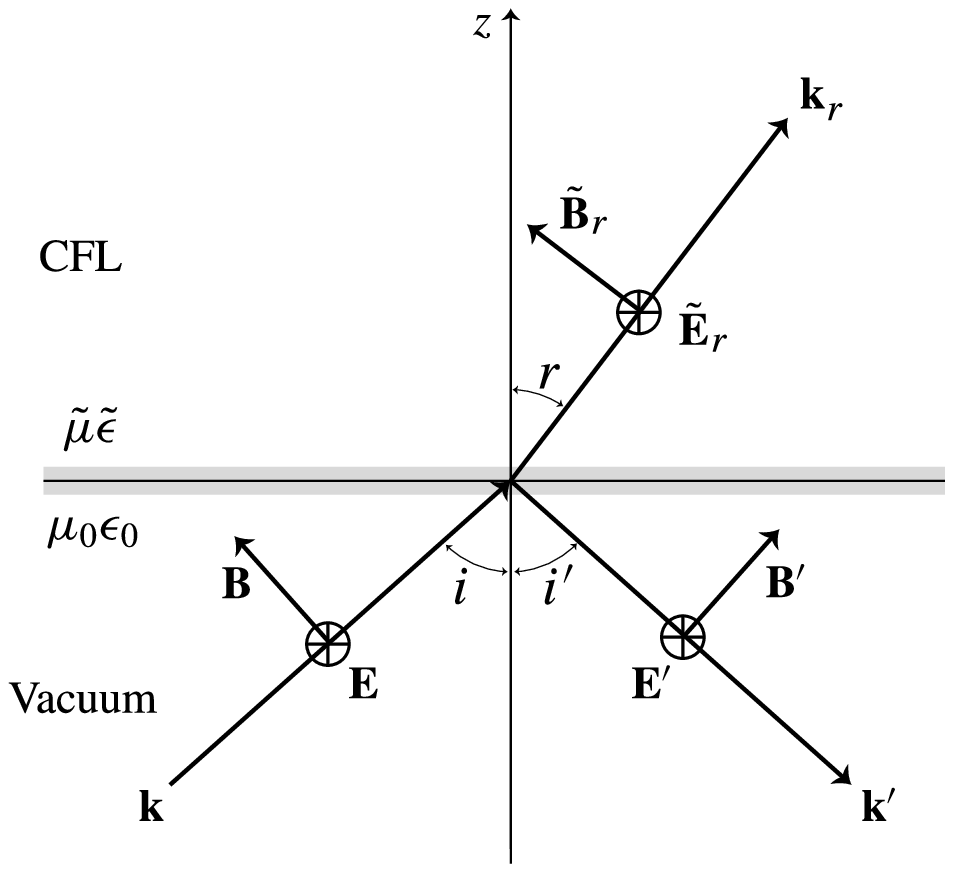}\hfill
\vspace{0.1in}
\epsfxsize=2.7in
\hspace*{0.233in}
\epsffile{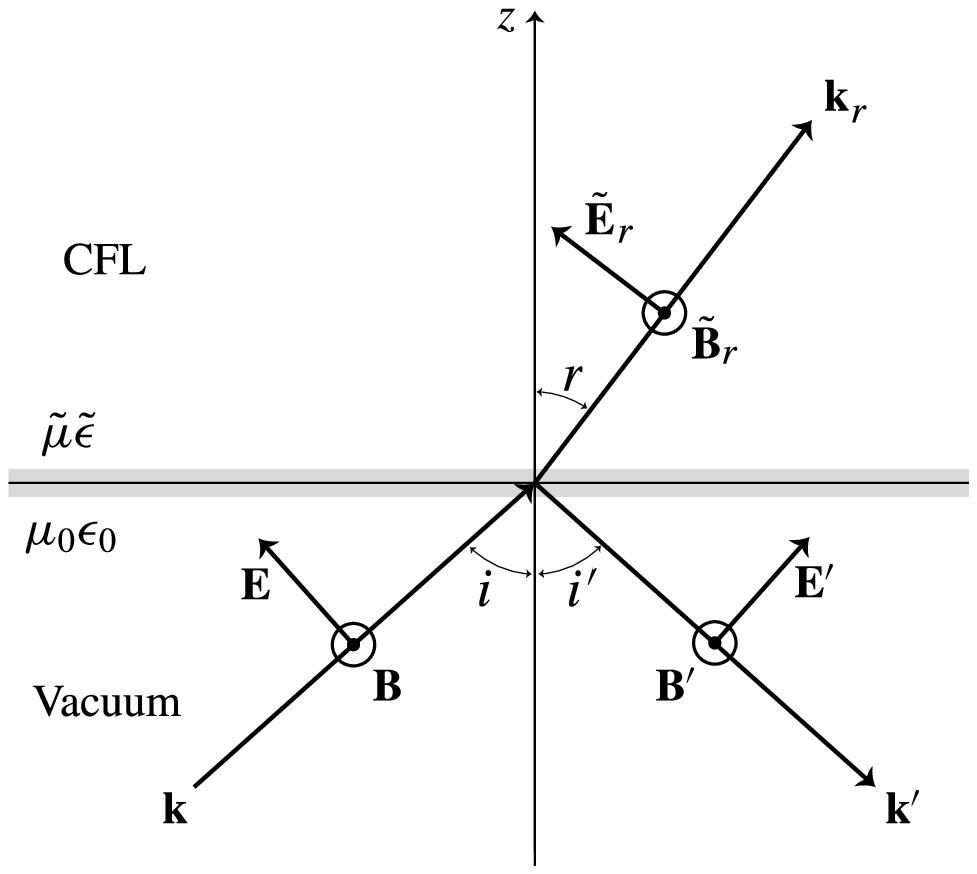}
\label{fig1}
\vspace{0.1in}
\caption{Incident wave $\bf k$ strikes a planar interface between
vacuum and CFL quark matter, giving a reflected wave
$\bf k'$ and a refracted wave ${\bf k_r}$.  We use boundary
conditions to relate electromagnetic fields just below
the interface to $\tilde Q$-electromagnetic fields just above,
assuming that the wavelength of the light is long compared to
the screening length in the CFL phase and the confinement
length in vacuum, symbolized by grey shading. 
Top panel: polarization perpendicular
to the plane of incidence and thus parallel to the interface.
Bottom panel: polarization parallel to
the plane of incidence.}
\vspace{-0.2in}
\end{figure}

The boundary conditions must be obeyed at all times, which immediately
implies that the frequency of all the waves is the same, as above.
The boundary conditions must be obeyed at all points on the
planar interface.  For $1/k\gg \lambda^{\rm CFL},\lambda^{\rm QCD}$
this implies that
${\bf k}\cdot {\bf r}={\bf k'}\cdot {\bf r}={\bf k_r}\cdot {\bf r}$ 
at $z=0$, independent of details of the boundary conditions. 
To satisfy this kinematic constraint, all three wave
vectors must lie in a plane and 
$k \sin{i} = k' \sin{i'} = k_r \sin{r}$.
Since $k = k'$, we must have $i = i'$:
that is, the angle of incidence is the same as the angle of reflection.
Since $k_r= \tilde n k$, we also reproduce Snell's law
\begin{equation}
\label{Snell}
\sin{i} = {\tilde n} \sin{r} \ .
\end{equation}
The kinematics of the reflection and refraction of light
on CFL quark matter 
are unaffected by the mixing angle $\theta$.

We now use the boundary conditions 
to find the intensities of the reflected and refracted radiation.
For the first polarization in Fig.~1, (\ref{ParallelH}) and (\ref{ParallelE})
yield
\begin{eqnarray}
\cos\theta\,({\mathcal{E}}-{\mathcal{E'}})\sqrt{\frac{\epsilon_0}{\mu_0}}
\cos i &=&
\tilde{\mathcal{E}}_r \sqrt{\frac{\tilde\epsilon}{\tilde\mu}} \cos r \ ,
\nonumber\\
\mathcal{E}+\mathcal{E'}&=&\cos\theta\, \tilde{\mathcal{E}}_r\ ,
\end{eqnarray}
and, using Snell's law, 
(\ref{PerpendicularB}) is equivalent to (\ref{ParallelE}) in this case. 
Solving, we find
\begin{mathletters}
\label{pol-perpen}
\begin{eqnarray}
\ & \ & \ \nonumber\\ 
\frac{\tilde{\mathcal{E}}_r}{\mathcal{E}} 
&  = & \frac{2 \cos{\theta}\, \cos{i}}{
\cos^2{\theta}\, \cos{i} + \frac{\mu_0}{{\tilde \mu}}\,\tilde n \cos r 
} 
\ , \\
\frac{ \mathcal{E}'}{\mathcal{E}} 
&  = & \frac{ \cos^2{\theta}\, \cos{i}
-\frac{\mu_0}{{\tilde \mu}} \,\tilde n \cos r
} 
{\cos^2{\theta}\, \cos{i} + \frac{\mu_0}{{\tilde \mu}}\, \tilde n \cos r
} 
\ ,
\end{eqnarray}
\end{mathletters}
where $r$ is easily eliminated using Snell's law in the form
$\tilde n \cos r = \sqrt{\tilde n^2-\sin^2 i}$.
To the order we are working, ${\tilde \mu}=\mu_0$. 
For the second polarization of Fig.~1, (\ref{ParallelE})
and either (\ref{ParallelH}) or (\ref{PerpendicularD}) yield
\begin{eqnarray}
({\mathcal{E}}-{\mathcal{E'}})\cos i &  = &  \cos\theta \,
\tilde{\mathcal{E}}_r \cos r  \ , \qquad
\nonumber\\
\cos\theta\,\sqrt{\frac{\epsilon_0}{\mu_0}}\,(\mathcal{E}+\mathcal{E'})
&  = &  
\sqrt{\frac{\tilde\epsilon}{\tilde\mu}} \tilde{\mathcal{E}}_r\ , 
\end{eqnarray}
and hence
\begin{mathletters}
\label{pol-parallel}
\begin{eqnarray}
\frac{\tilde{\mathcal{E}}_r}{\mathcal{E}} 
&  = & \frac{2 {\tilde n} \cos{\theta} \cos{i}}
{ \frac{\mu_0}{{\tilde \mu}}\,   {\tilde n}^2 \cos{i} 
+ \tilde n \cos r\cos^2{\theta} 
} \ , \\
\frac{ \mathcal{E}'}{\mathcal{E}} 
&  = & \frac { \frac{\mu_0}{{\tilde \mu}} \,  
{\tilde n}^2 \cos{i} - \tilde n \cos r \cos^2{\theta} 
}
{ \frac{\mu_0}{{\tilde \mu}}  \, {\tilde n}^2 \cos{i} + \tilde n \cos r
\cos^2{\theta} 
} \ .
\end{eqnarray}
\end{mathletters}
Upon setting $\cos\theta=1$, the relations (\ref{pol-perpen})
and (\ref{pol-parallel}) 
reproduce results for 
reflection and refraction off standard dielectric media 
(see Ref. \cite{Jackson}).  
Decreasing $\cos\theta$ decreases the $A_\mu$ component of 
$A^{\tilde Q}_\mu$, and thus 
favors reflection over refraction.
For $\theta$ as small as in nature, the changes
introduced by $\theta\neq 0$ are small.  
In the (unphysical) limit in which 
$\cos \theta=0$, $A^{\tilde Q}_\mu$ 
would be orthogonal to $A_\mu$ making the CFL phase
a superconductor with respect to ordinary
electromagnetism.  In this limit, we expect and find
zero refraction and perfect reflection for both polarizations.

The value of Brewster's angle is modified by a non-vanishing $\theta$.
This is the incident angle for which there is no reflected wave
if the polarization is parallel to the plane of incidence.
We find
\begin{equation}
\label{Brewster}
i_B = \arctan{\left(
\frac{{\tilde n}}{\cos^2{\theta}}
 \sqrt{\frac{{\tilde n}^2-\cos^4{\theta}  }{{\tilde n}^2-1}} \right) } \ .
\end{equation}
We can also imagine sending a ${\tilde Q}$ electromagnetic wave from 
CFL matter into vacuum. Doing this calculation yields the same 
results, but with $\tilde n$ replaced by $1/\tilde n$. 
As usual, Snell's law then implies that for incident angles bigger than
$\arcsin{(1/\tilde n)}$, 
the ${\tilde Q}$-light cannot escape from the  CFL matter.

Are the solutions (\ref{pol-perpen}) and (\ref{pol-parallel}) 
consistent with energy conservation? 
The Poynting vector ${\bf S} = \frac 12 \left( {\bf E} \times {\bf H} \right)$
measures the energy flow per unit area and time. Continuity of
the $z$-component
of the Poynting vector requires
\begin{equation}
\label{energy-con}
{\mathcal{E}}^2 \sqrt{\frac{\epsilon_0}{\mu_0}}\cos i 
= ({\mathcal{E}}')^2 \sqrt{\frac{\epsilon_0}{\mu_0}}\cos i 
+ (\tilde{\mathcal{E}}_r)^2 
\sqrt{\frac{\tilde\epsilon}{\tilde\mu}}\cos r\ ,
\end{equation}
a relation which is indeed satisfied by both 
(\ref{pol-perpen}) and (\ref{pol-parallel}).

Notice that in our analysis of each of the two polarizations,
one boundary condition was irrelevant and Snell's law could
be used to eliminate a second.  If we had
used energy conservation in the form (\ref{energy-con}) in
our derivation instead of just as a check, we could have derived
all our results from the single boundary condition (\ref{ParallelH}).
That is, given only the
boundary condition
(\ref{ParallelH}) which is easily derived, 
Snell's law (\ref{Snell}) which is kinematic, 
and energy conservation (\ref{energy-con}),
we can derive our solutions describing the reflection and refraction of light
of both polarizations and, from these electromagnetic fields, 
we can then derive the remaining boundary
conditions (\ref{PerpendicularD}), (\ref{PerpendicularB})
and (\ref{ParallelE}). This means 
that we have {\it derived} the boundary
conditions motivated above by the idea that the
QCD vacuum behaves like a dual superconductor filled with a condensate of 
color-magnetic monopoles~\cite{tHooft}.
Having analyzed the illumination of dense quark matter, we find
that in addition we have illuminated our understanding of the QCD vacuum.
\\[-1.5ex]

We thank D. Son for a stimulating conversation. CM thanks the MIT
CTP for hospitality during an early stage of this work.
The work of CM is supported by the European Community through 
Marie Curie Grant \#HPMF-CT-1999-00391. The work of KR  
is supported in part  by the 
DOE under agreement \#DF-FC02-94ER40818 and an OJI grant,
and by the A. P. Sloan Foundation. 
\\[-3.4ex]

\end{document}